\documentclass[11pt,twoside]{article}
\usepackage{asp2006}
\usepackage{graphics}
\usepackage{epsf}
\usepackage{lscape}
\def\edcomment#1{\iffalse\marginpar{\raggedright\sl#1\/}\else\relax\fi}
\reversemarginpar

\begin{document}
\title{Evolution of a Coronal Twisted Flux Rope}
\author{Nour-Eddine Raouafi}
\affil{Johns Hopkins University Applied Physics Laboratory, Laurel, MD, USA. e-mail: Nour.Eddine.Raouafi@jhuapl.edu}

\begin{abstract}
Multi-instrument observations of NOAA AR10938 on Jan. 14-18, 2007, are utilized to study the evolution of a magnetic thread system with multiple crossings suggestive of a twisted coronal flux rope. A C-class flare recorded by GOES on Jan. 16, at approximately 2:35~UT led to the brightening of the structure, that is seen in Hinode/EIS data at 2:46~UT, Hinode/XRT after 2:50~UT, and {\emph{STEREO}}/SECCHI/EUVI images at 3:30~UT. 304~{\AA} images revealed the presence of rapidly evolving, dark fibrils along the bright structure before and after the flare. A denser structure formed a few hours later and lasted for several days forming a segment of an inverse S-shaped filament. The present set of data is highly suggestive of the presence of a twisted flux rope prior to the formation of the filament segment at the same location.
\end{abstract}

\vspace{-1cm}
\section{Introduction}

Multi-wavelength observations show that most solar eruptive events (flares and coronal mass ejections: CMEs) are associated with magnetic fields with complex topologies \citep[twist, shear, writhe, and linking; see][]{Canfield99}. Hence, numerous recent studies were dedicated to characterize the formation of solar filaments and their eruptive evolution into CMEs. Resolving the magnetic structure of these phenomena is therefore important to constrain models \citep[see][]{Chae00}. 

We use multi-instrument observations of NOAA AR10938 on Jan. 14-18, 2007, to study the evolution of a magnetic thread system with multiple crossing that is highly suggestive of a flux rope.

\vspace{-0.5cm}
\section{Observations and Data Analysis}

\citet{Raouafi09} reported on a non-eruptive C-class flare recorded by GOES in AR10938 on Jan. 16, 2007, at approximately 2:35 UT, which led to the brightening of the complex magnetic thread system. Data from the {\emph{Hinode}} X-Ray telescope \citep[XRT:][]{Golub07}, the Extreme UV Imaging Spectrometer \citep[EIS:][]{Culhane07} and the Solar Optical Telescope \citep[SOT:][]{Tsuneta08} filtergram are used to study the formation and evolution of the structure along with EUV images from {\it{STEREO}}/SECCHI/EUVI \citep[EUVI: ][]{Howard08}. The different data sets provide a temperature coverage ranging from $\sim0.08$~MK to $>10$~MK. This is important to study the temporal evolution of the structure.

\begin{figure*}[!h]
\plotone{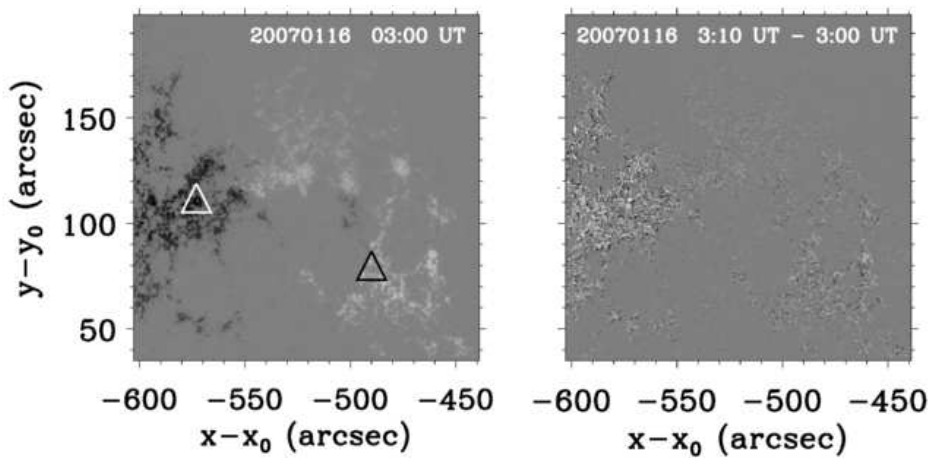} 
\caption{Left: LOS-magnetogram from {\emph{Hinode}}/SOT-FG. The triangles indicate approximately the footpoint locations of the magnetic loop system. Right: difference of the unsigned magnetic flux. \label{hinode_sotfg}}
\end{figure*}

\begin{figure*}[!h]
\plotone{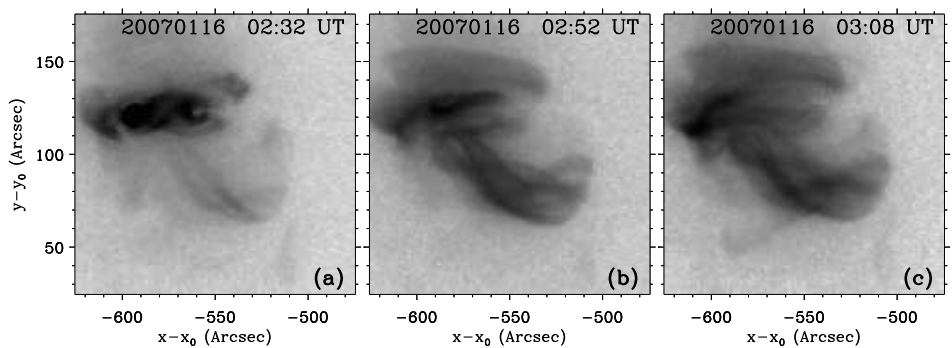}
\caption{ NOAA AR10938 as seen by {\emph{Hinode}}/XRT prior to and after the flare. The magnetic thread system is best seen around 3:00~UT. The individual threads indicate a multiple crossing topology highly suggestive of a flux rope. \label{hinodexrt}}
\end{figure*}

Fig.~\ref{hinode_sotfg} displays a LOS-photospheric magnetogram from SOT-FG of the active region. The magnetic structure extends between the triangle symbols that mark approximately the locations of its footpoints. The loop-system ends are rooted into pores and plage areas with opposite dominant polarities, where relatively important changes of the flux were occurring (see right panel of Fig.~\ref{hinode_sotfg}). The GOES non-eruptive, C-class flare took place near the northeast footpoint (see Fig.~\ref{hinodexrt}a). 

The full extend of the magnetic thread system is revealed by emissions of high temperature spectral lines observed by EIS (Fe~{\sc{xxiv}} 255.1~{\AA}: $\log T=7.2$;  Ca~{\sc{xvii}} 192.82~{\AA}: $\log T=6.7$) at about 2:46 UT. The loop topology in EIS data is similar to the X-Ray one recorded about 15-20 minutes after the flare (Fig.~\ref{hinodexrt}) with, however, a better contrast. It is noteworthy that emissions in low temperature lines did not show similar features. X-Ray data also show that the structure had an apparent simpler topology prior to the flare.

\begin{figure*}
\plotone{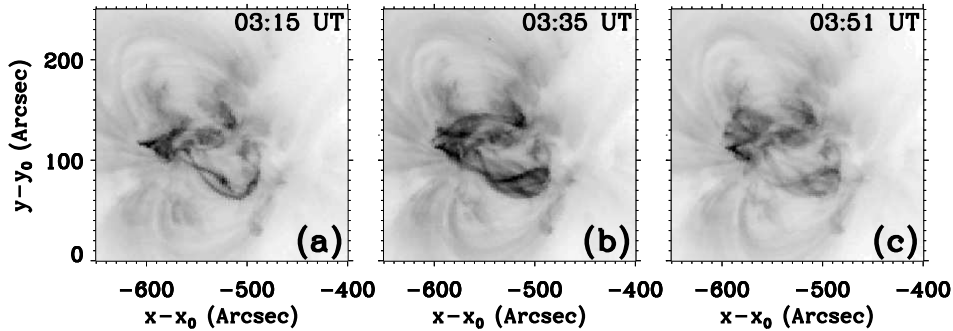}
\caption{195~{\AA} images from EUVI-A on Jan. 16, 2007, illustrating the evolution of the EUV counterpart of the X-ray threads observed by {\emph{Hinode}}/XRT. The EUV structure looks similar to that observed in X-rays with, however, a time gap greater than 30 minutes in appearance. \label{secchi_euvia}}
\end{figure*}

EUVI-A observations in the 171~{\AA} and 195~{\AA} channels show the relatively cool ($\sim1$~MK) counterpart of the loop system observed earlier by EIS and XRT. Fig.~\ref{secchi_euvia} displays snapshots of AR10938 recorded between 3:00 UT and 4:00 UT. The EUVI structure is very similar to the one observed earlier in EIS's hot lines with, however, a delay greater than 45 minutes. The arrangement of the threads is better seen around 3:30 UT and show similarities with EIS and XRT observations.

The sequential appearances of the suggested flux rope in hot emission lines (e.g., Fe~{\sc{xxiv}} 225.1~{\AA}, $\log T=7.2$), then in X-ray images (a few MK), and finally in EUV 171~{\AA} and 195~{\AA} images ($\sim1$~MK) show different cooling stages of the magnetic structure and display details of its fine structure.

\begin{figure*}[!h]
\plotone{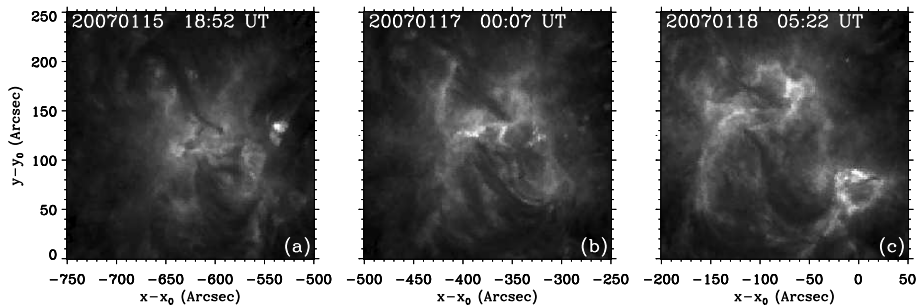}
\caption{The filament along the neutral line of
AR10938 is well developed north the active region and is of filamentary nature south of it. (b-f) 304~{\AA} images from {\it{STEREO}}/SECCHI/EUVI-A illustrating the evolution of the filament segment corresponding  to the thread system seen in EUV and X-rays.\label{secchi_euvia_304}}
\end{figure*}

EUVI observations of cooler material in the He~{\sc{ii}} 304~{\AA} line reveal the presence of dark fibrils along the path of the magnetic thread structure observed in EUV and X-ray data. These fibrils were present several hours before the manifestation of the indicated flux rope (see Fig.~\ref{secchi_euvia_304}a). Their appearance was changing rapidly with time. These are probably part of a larger dense structure forming an inverse S-shape (see Fig.~\ref{secchi_euvia_304}a-b) extending along the neutral line across AR10938 (Fig.~\ref{secchi_euvia_304}). The presence of the filament was more prominent in the following few days, in particular on Jan. 18, 2007, (see Fig.~\ref{secchi_euvia_304}b-c). We believe this is a segment of the larger, inverse S-shaped filament structure. This supports the hypothesis that the topology of the different threads indicates a high degree of linking suggesting the presence of a twisted flux rope (see Fig.~\ref{secchi_euvia}b-c).

\section{Conclusions and Discussion}

The presence of dark fibrils indicates that magnetic structure may have been present prior to the GOES C-class flare. These were, however, of filamentary nature and evolving rapidly with time. It is likely that the flux tube emerged twisted from the convection zone. Photospheric shear motions and helicity transfer during the flare may also contribute to the twist of the flux tube. Since the flare is non-eruptive, the helicity of the flaring system is likely to be evacuated to neighboring flux tube by increasing their topological complexity.

The presence of helical structures during filament eruptions is well established by observations. However, their role in the formation of solar filaments and also their occurrence prior to the eruptive phase of CMEs are still highly debated. Two classes of models are proposed to address these questions. The first class is flux rope based models, which consider highly twisted flux tubes to be at the origin of solar filaments \citep{RustKumar94,PriestForbes90,Low01}. The second class considers highly sheared arcades along magnetic inversion lines to be the base of the filament, where the helical structure occurs only during the eruptive phase of the filament and that is the result of the ongoing magnetic reconnection \citep{Pneuman83,vanBallegMart89,vanBallegMart90,Antiochos94}. Although the present observations are not meant to discriminate between the two types of models, they indicate the presence of a flux rope prior to the filament formation that lasted for several days. We believe that it favors the flux-rope based models. For more details see \citet{Raouafi09}.


\begin{thebibliography}{}

\bibitem[Antiochos et al.(1994)]{Antiochos94} Antiochos, S. K., Dahlburg, R. B., \&
Klimchuk, J. A. 1994, \apj, 420, L41

\bibitem[Canfield et al.(1999)]{Canfield99} Canfield, R. C., Hudson, H. S., \& McKenzie, D. E. 1999, GRL, 26, 627

\bibitem[Chae(2000)]{Chae00} Chae, J. 2000, \apj, 540, L115

\bibitem[Culhane et al.(2007)]{Culhane07} Culhane, J.L., Harra, L.K, James, A.M., et al. 2007, \solphys, 243, pp. 19-61

\bibitem[Golub et al.(2007)]{Golub07} Golub, L., et al. 2007, \solphys, 243, 63

\bibitem[Howard et al.(2008)]{Howard08} Howard, R. A., et al. 2008, \ssr, 136, 67

\bibitem[Low(2001)]{Low01} Low, B. C. 2001, JGR, 106, 25141

\bibitem[Pneuman(1983)]{Pneuman83} Pneuman, G. W. 1983, \solphys,88, 219

\bibitem[Priest \& Forbes(1990)]{PriestForbes90} Priest, E. R., \& Forbes, T. G. 1990, \solphys,
126, 319

\bibitem[Raouafi(2009)]{Raouafi09} Raouafi, N.-E. 2009, \apj, 691, L128

\bibitem[Rust \& Kumar(1994)]{RustKumar94} Rust, D. M., \& Kumar, A. 1994, \solphys, 155, 69

\bibitem[van Ballegooijen \& Martens(1989)]{vanBallegMart89} van Ballegooijen, A. A., \& Martens, P.
C. H. 1989, \apj, 343, 971

\bibitem[van Ballegooijen \& Martens(1990)]{vanBallegMart90} van Ballegooijen, A. A., \& Martens, P.
C. H. 1990, \apj, 361, 283

\bibitem[Tsuneta et al.(2008)]{Tsuneta08} Tsuneta, S., Suematsu, Y., Ichimoto, K., et al. 2008,
\solphys, 249, 167

\end{thebibliography}
\end{document}